\documentstyle[12pt,epsfig]{article}
\let\section=\subsection     \let\subsection=\subsubsection                

\begin{document}
\begin{center}
{\large \bf RECENT RESULTS FROM E866 AT BNL}\footnote{presented at
the International Workshop XXV on Cross Properties of Nuclei and
Nuclear Excitations, Hirschegg, Austria, January 1997}\\[5mm]
CHRISTIAN M\"UNTZ\\[3mm]
{\small \it  Department of Physics, Brookhaven National Lab.,
Upton, NY 11973, U.S.A.}\\[2mm] 
for Experiment 866 (E802 Collaboration):\\
{\small \it BNL-UC(Space Science Lab.)-UC(Riverside)-Coulumbia-\\
Tokyo-INS-Kyoto-LLNL-MIT-Tsukuba-Yonsei-Maryland}
\end{center}

\begin{abstract}\noindent
Recent and preliminary single-particle data are presented from the AGS
Experiment~866 at BNL. Emphasis is put on
the transverse mass as well as
the rapidity distributions of charged pions, kaons, protons, deuterons and
anti-protons measured in the most central Au+Au collisions at an incident
kinetic energy of 10.8~AGeV. The data suggest a high degree of
stopping power present in these reactions. Applying an expanding
fireball scenario to describe the experimental distributions 
substantial transverse and longitudinal flow velocities result.
\end{abstract}

\vspace*{-0.4cm}
\section{Introduction}
Heavy ion collisions at (ultra-) relativistic energies 
provide an unique tool to study the properties of nuclear matter far
from its groundstate. New and exotic physics is expected to take place
in violent collisions of very heavy ions like Au nuclei. These
reactions are presently under study covering a huge range of incident
energy. Experiment~866, installed at AGS/BNL, was specially designed to
investigate Au+Au reactions in the energy regime of 10~AGeV~\cite{E866}. The 
E866 spectrometers provide particle identification
and momentum measurement for a variety of charged particles. 
Central reactions can be selected by means of an event characterization.
In addition, a program of two-particle 
correlation measurements is pursued.
Having briefly introduced the experimental setup,
this contribution emphasizes on the recent and preliminary
single-particle data of $\pi^\pm$, $K{}^\pm$, $p$, $\bar{p}$ and
$d$ from the most violent Au+Au reactions at 10.8~AGeV incident
kinetic beam energy. These data will be discussed
in terms of transverse and longitudinal flow.

\vspace*{-0.2cm}
\section{Experimental Setup}
The experiment consists of three parts. 
First, a combination of several beam counters located 
before and after the target is used to monitor the beam quality 
and to provide the minimum bias and interaction trigger.
Second, 
several global detectors provide the event characterization.
Here, the zero-degree calorimeter was used to
identify the most central collisions by requiring a minimal energy
deposition of projectile nucleons
in a cone around the beam axis with an opening angle of 1.5~degrees.  
The software cut to identify the most central collisions corresponds
typically to trigger cross sections 
below 10\% of the interaction cross section. 
Third, two separate magnetic spectrometers for particle
identification and momentum measurement are used: 
the Henry Higgins (HH) and the Forward Spectrometer
(FS)~\cite{SPECTRO}. Both are equipped with
sophisticated time-of-flight and tracking capabilities.
The small solid angle of FS of 5~msr, compared to 25~msr of HH, allows
to handle the high multiplicities at
emission angles as small as 6~degrees in the laboratory.
E866 achieves an optimum and unique phase space coverage by
the interplay of both spectrometers.
As a consequence rapidity distributions can be computed within
$y=1.6 \pm 1$ and even beyond depending on the
particle species, taking advantage of the phase space symmetry in
mass-symmetric reactions.

\vspace*{-0.3cm}
\section{Preliminary Data from Central Au+Au Reactions}
\begin{figure}[t]
\begin{center}
\begin{minipage}[h]{14cm}
\begin{center}
\mbox{\epsfig{width=14cm,height=9.0cm,
file=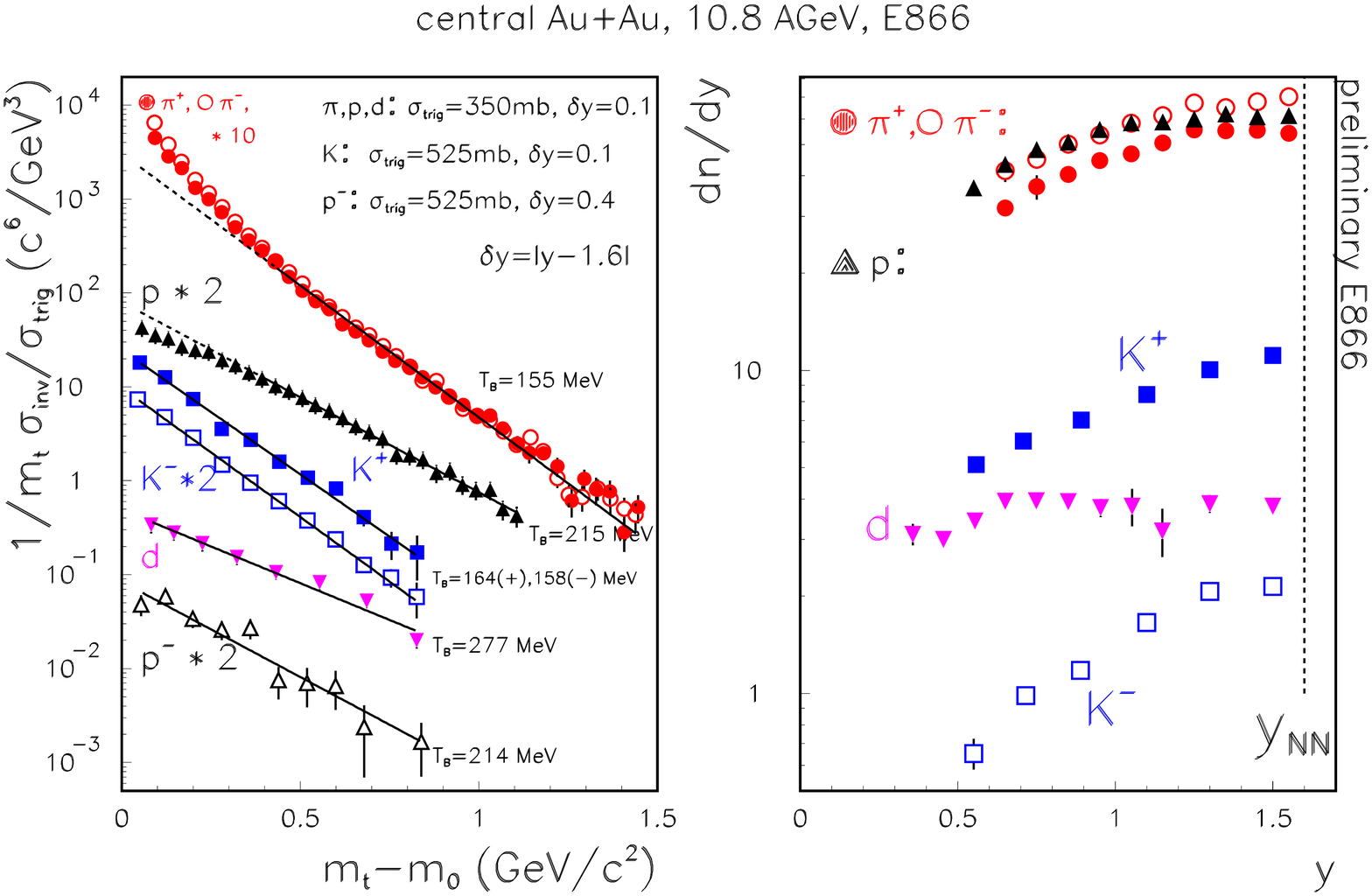
}}
\vspace*{-0.7cm}
\caption{
{\it
Left:
Spectra at midrapidity measured in the
most central reactions, plotted in a Boltzmann representation.
Inverse slope parameters~$T_B$ are indicated, resulting from fitting
Boltzman distributions to the spectra (solid lines).
$\sigma_{inv}=d^2\sigma / (2\pi m_t dm_t dy)$.
Right:
Corresponding rapidity distributions,
shown in one hemisphere only. Statistical errors only, 
systematic errors estimated to be $\pm(10-15)$\%.
Depending on phase space coverage, data points are combined 
taking advantage of the symmetry around midrapidity. 
}
}\label{DATA}
\end{center}
\end{minipage}
\end{center}
\vspace*{-0.9cm}
\end{figure}

All data presented here are taken from ref.~\cite{E866}.
Figure~\ref{DATA}, left,  shows the kinetic transverse mass
spectra at midrapidity for
different particle species for the most central Au+Au collisions 
at 10.8~AGeV incident kinetic energy. 
The corresponding apparent temperatures (inverse slope
parameters) are indicated. They range from 150~($\pi$) to
280~MeV~(d) and increase with
increasing rest mass of the particle.
Note, that the spectra of protons and charged pions significantly
deviate from a Boltzmann distribution
and a lower fit limit of 0.4~GeV/c${}^2$ in transverse kinetic energy
was used. In addition,
negative pions are more enhanced at low energies than positive pions.
This indicates the influence of Coulomb effects caused by the charged
nuclear matter. A quantitative
analysis of this effect based on a static Coulomb source located at
midrapidity is consistent with an overall ratio
$\pi^-/\pi^+\approx 1.2$ and an effective Coulomb boost of about $\pm 9$~MeV.

Figure~\ref{DATA}, right, shows the rapidity distributions after
integrating the spectra over transverse mass. For pions,
protons and deuterons fit functions based on the sum of two 
exponentials have been used.
The statistics for anti-protons is not sufficient to compute
a rapidity distribution. One observes for central collisions that all
distributions exhibit a maximum at midrapidity. The width of the
distributions depends on the particle species and it is larger for
protons and deuterons than for produced particles. 
The experimental ratio of negative to positive pions of
1.25$\pm$0.06 is almost independent from rapidity. 
The distinct broad maximum of the proton rapidity distribution
suggests a high degree of stopping in central Au+Au collisions at
this beam energy regime. 
This interpretation is supported by two observations: 
This maximum gradually changes to a
pronounced minimum at midrapidity when reducing 
(i) the violence of the collision, or 
(ii) the combined mass of the projectile and target,
compare to  Si+Al~\cite{SIAL} and p+p~\cite{PP} reactions. 

\vspace*{-0.2cm}
\section{Transverse and Longitudinal Flow}
\begin{figure}[t]
\begin{center}
\begin{minipage}[h]{14cm}
\begin{center}
\mbox{\epsfig{width=14.0cm,height=9.0cm,
file=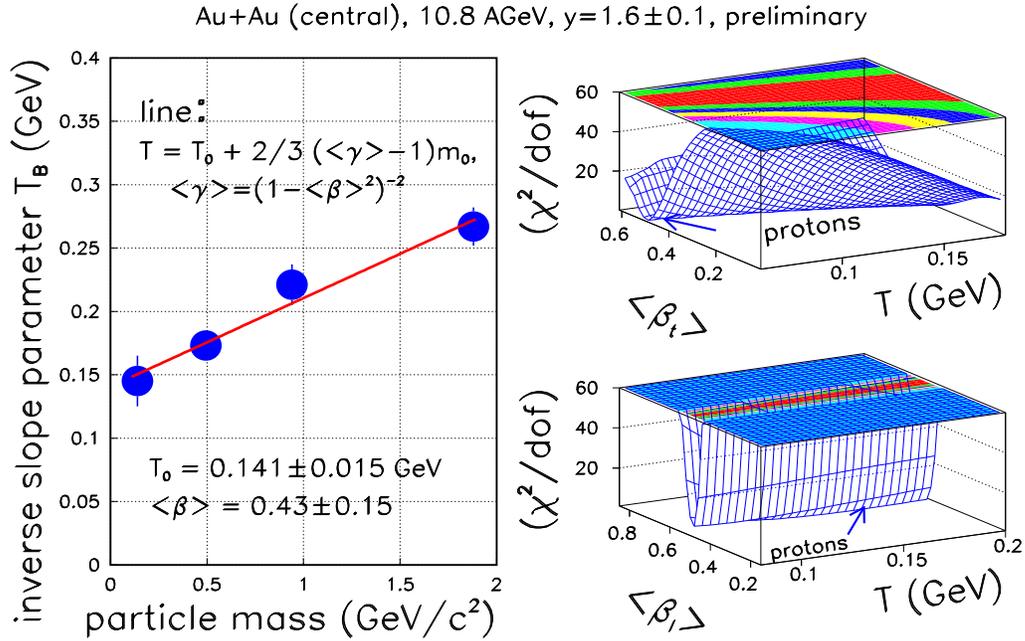
}}
\vspace*{-0.9cm}
\caption{
{\it
Left:
Apparent inverse slope parameters resulting from fitting Boltzmann
distributions to the transverse mass spectra,
as a function of the particle rest mass.  
Right: 
Proton $\chi^2/dof$ surfaces, plotted as a function of the corresponding
temperature and flow velocity parameter sets: 
from transverse mass spectra (upper part) and 
rapidity distributions (lower part). 
Arrows indicate the absolute minima.
}
}\label{TEMP}
\end{center}
\end{minipage}
\end{center}
\vspace*{-0.7cm}
\end{figure}

The left part of fig.~\ref{TEMP} summarizes the 
inverse slope parameters at midrapidity 
for different particle species, as
a result from fitting Boltzmann distributions to the asymptotic
high-energy part of the spectra, see fig.~\ref{DATA}. The apparent
temperatures exhibit a linear dependence on the particle rest mass, as
it has been also observed at lower~\cite{EOS} and higher~\cite{NA44}
incident energies. 
One may assume that the apparent 
temperature is proportional to the
sum of the average thermal and radial flow kinetic energy. 
Based on that
the fit shown in fig.~\ref{TEMP} gives an estimation of
the temperature of about 140~MeV and the flow velocity of about 0.4.
These results are very similar to what has been analysed for Si+Au
data in the same energy regime~\cite{PBM}.
In the following a
more sophisticated fitting procedure  
is taken to deduce the common parameter set
temperature~$T$, average transverse and longitudinal flow velocities
$<\beta_t>$, $<\beta_l>$, respectively, from the measured transverse
mass spectra and rapidity distributions.
This method is described in 
ref.~\cite{SCHNED} and has been also used in ref.~\cite{NA44,PBM}.
These parameters characterize
the reaction at freeze-out. The formalism describes an exploding
fireball with an cylindrical rather than a spherical expansion profile
to account for the possibility of a forward-backward asymmetry. 
It is assumed that the transverse
and longitudinal motion of the thermal source are decoupled.

The procedure is applied to pions, positive kaons, protons and deuterons.
The strategy is to explore the parameter space for each particle
species in the transverse and the longitudinal degree of freedom,
i.e.~transverse mass spectra and rapidity distributions, independently,
and to compare the final $\chi^2$-surfaces. In contrast to ref.~\cite{PBM}
the temperature is not deduced from experimental particle ratios,
but results from the fitting procedure.
The transverse velocity profile was chosen to depend linearly on the
radius and its average value results from integration over the profile,
assuming a freeze-out radius of 6.5~fm. Within the given errors 
the fit results do not significantly
depend on the choice of the freeze-out radius and the velocity
profile. A lower fit limit of $m_t-m_0=0.4$~GeV/c${}^2$ was introduced
for the pion spectrum to minimize the influence of resonance
contributions on the results.
The average longitudinal velocity is deduced from the
limits of the rapidity interval within which the individual thermal
sources are distributed and superimposed.

The right part of fig.~\ref{TEMP} shows the corresponding $\chi^2$
surfaces for fitting the proton transverse mass spectrum (upper part) and
rapidity distribution (lower part). The arrows indicate the absolute minima
which are not very pronounced along the valleys.
Temperature and transverse flow velocity are anti-correlated, whereas
temperature and longitudinal flow are almost decoupled.
In order to find a common parameter set, the corresponding projections
of the $\chi^2$ surfaces of the different particle species
are overlayed.
An overlapping region for all particles exists for
the transverse degrees of freedom, and hence, a common set of 
temperature and average transverse flow velocity can be deduced.  
The results for the longitudinal degrees of freedom are not that
promising. The maximum of the positive kaon rapidity distribution 
turns out to be
too pronounced in order to agree with a longitudinal flow scenario
with substantial flow. On the other hand,
the width of the deuteron distribution appears to be too large
to deduce a common velocity for deuterons and protons.
Besides from that the following common data set represents a fair compromise:
$
T = (127+10-15) MeV, \; 
<\beta_t> = 0.39\pm 0.05, \;
<\beta_l> = 0.50+0.10-0.05
$.
The error bars are deduced from the size of the overlap region
of the $\chi^2$-surfaces. 
\begin{figure}[t]
\begin{center}
\begin{minipage}[h]{14cm}
\begin{center}
\mbox{\epsfig{width=12.0cm,height=8.0cm,
file=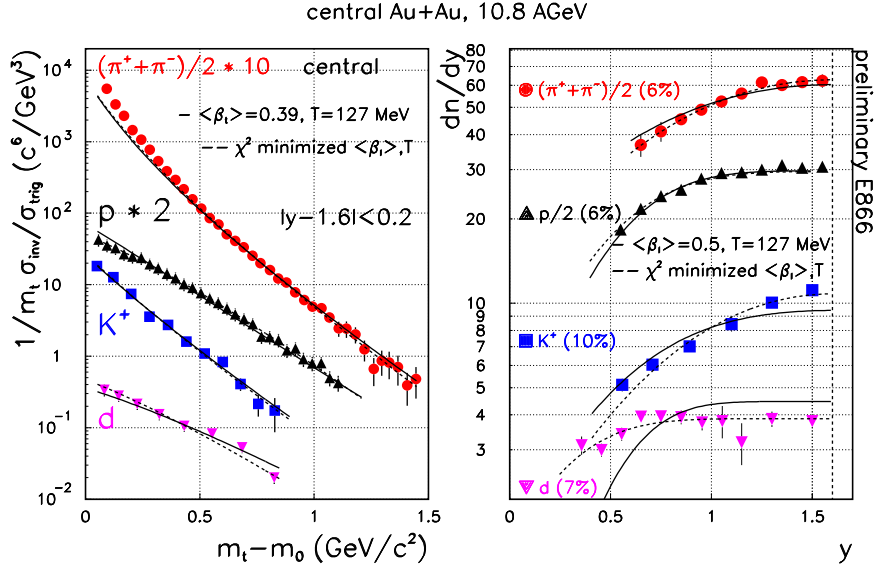
}}
\vspace*{-0.5cm}
\caption{
{\it
Comparison between data (symbols, see fig.~1 for details) 
and fit results (lines) for
the transverse (left) anf longitudinal (right) fitting procedure.
}
}\label{RESULTS}
\end{center}
\end{minipage}
\end{center}
\vspace*{-0.7cm}
\end{figure}
Figure~\ref{RESULTS} summarizes the fit results for the common
parameter set (solid lines) and the individual best fits (dashed lines).

\vspace*{-0.2cm}
\section{Summary}
Recent and preliminary data from E866 at AGS/BNL are presented. The
data from both E866
spectrometers have been successfully combined to cover a huge portion
of the phase space for a variety of particle species. Focusing on
midrapidity transverse mass spectra and rapidity distributions a flow
analysis was performed. The resulting temperature and average flow velocities
agree with the trend suggested by the corresponding analysis at
different energy regimes~\cite{EOS,NA44}.

\end{document}